\documentclass[nofootinbib,letterpaper,aps,amsmath,amsfonts,twocolumn]{revtex4}
\usepackage[mathscr]{eucal}
\newcommand{\td}{\textup{d}}  
\def\Pcm#1{{\mathcal{#1}}}    
\newcommand{\del}{\partial}   
\def\eqref#1{(\ref{#1})}      
\def\sref#1{[\ref{#1}]}       
\def\er#1{eqn.\eqref{#1}}     
\def\nn{\nonumber}
\begin{document}
\title{Covariant Calculus for Effective String Theories}
\author{N.D. Hari Dass}
\affiliation{Centre for High-Energy Physics, Indian Institute of Science, Bangalore, India\footnote{from September 26th}}
\email{dass@imsc.res.in}
\author{Peter Matlock} 
\affiliation{Department of Electrophysics, National Chiao Tung University, Hsinchu, Taiwan}
\email{pwm@mail.nctu.edu.tw}
\begin{abstract}
A covariant calculus for the construction of effective string theories
is developed.  Effective string theory, describing quantum string-like
excitations in arbitrary dimension, has in the past been constructed
using the principles of conformal field theory, but not in a
systematic way. Using the freedom of choice of field definition, a
particular field definition is made in a systematic way to allow an
explicit construction of effective string theories with manifest exact
conformal symmetry.  The impossibility of a manifestly invariant
description of the Polchinski-Strominger Lagrangian is demonstrated
and its meaning is explained.
\end{abstract}

\maketitle

\section{Introduction}
\label{intro}

Although fundamental string
theory is of course confined to certain critical dimensions,
string-like phenomena do indeed appear as defects, solitons or
effective descriptions in a variety of physical situations. Since
these situations generically are of non-critical dimension, an
effective theory of strings must exist in order to describe them.

Polchinski and Strominger (PS) proposed the construction of such a
theory in \cite{PS}. As in other constructions of effective theories,
the formulation is required to exhibit the correct symmetries, while
dropping such requirements as renormalisability and polynomial
lagrangian, which are usually taken as minimal for a `fundamental'
theory expected to be valid at all energy scales. In particular, PS
treated an expansion around a long-string vacuum, where the
characteristic string length $R$ is taken as a large parameter. The
effective action is thus expanded in inverse powers of $R$.  The
notable difference with fundamental string theory is that the
effective PS theory contains a variable central charge, which can be
adjusted for consistency in any dimension.  Although Polchinski and
Strominger showed that the price one has to pay for this quantum
consistency in any dimension is the allowance of nonpolynomial terms
in the action, in such a perturbative expansion around the long-string
vacuum, such terms are perfectly acceptable.

PS were able to calculate the excitation spectrum including in the
effective action the first correction after the leading Polyakov-type
(equivalently Nambu-Goto) term. Surprisingly, the spectrum does not
deviate from that of Nambu-Goto theory at this order.  It has been
shown in \cite{Drum} and \cite{HDPM2} using an action valid to order
$R^{-3}$, that at even the next relevant order after this, the
spectrum does not differ from that of Nambu-Goto theory.

In the original formulation \cite{PS} of PS, the choice was made to
omit terms in the effective action proportional to the leading-order
equations of motion (EOM), which may be removed to appropriate order
by a field redefinition. In fact, as we have shown in \cite{HDPM2},
dropping or including a particular set of such `irrelevant
terms'\footnote{We retain the terminology used in \cite{PS,HDPM2},
  whereby terms proportional to the leading-order EOM are called
  \emph{irrelevant}, and terms irreducible to this form are deemed
  \emph{relevant}.} amounts to a particular choice of field
definition; the PS field definition corresponds to the omission of all
EOM terms.  It was pointed out in \cite{HDPM2} that different such
choices of field definition will correspond to actions invariant under
different transformation laws; different field definitions are related
by some field redefinition transformation, and this of course relates
potentially different transformation laws, each representing the
conformal symmetry of the theory.

The effective action proposed by Polchinski and Strominger is
\begin{eqnarray}
\label{PSaction}
S_{PS} &=& \frac{1}{4\pi} \int \td\tau^+ \td\tau^- \bigg\{
 \frac{1}{a^2} \del_+ X^\mu \del_- X_\mu \nn\\
&&\mbox{}+\beta \frac{\del_+^2 X\cdot\del_- X \del_+ X\cdot\del_-^2 X}{(\del_+X\cdot\del_-X)^2} 
+\Pcm{O}(R^{-3})
\bigg\}
.\end{eqnarray}
The quantity $R$ signifies the length of the string and in what follows, is taken to be large.
Consideration is restricted to fluctuations around the classical background.
The leading-order equation of motion
$
\del_+\del_- X^\mu = 0
$
has the solution
$
X^\mu_{\textup{cl}} = e^\mu_+R\tau^+ + e^\mu_- R \tau^-,
$
where $e_-^2=e_+^2=0$ and $e_+\cdot e_- = -1/2$.  

The action of \er{PSaction} is invariant, to the appropriate
order, under the transformation
\begin{equation}
\label{modtrans}
\delta^{\textup{PS}}_- X^\mu = \epsilon^-(\tau^-)\del_- X^\mu - \frac{\beta a^2}{2}\del_-^2 \epsilon^-(\tau^-)
\frac{\del_+ X^\mu}{\del_+X\cdot\del_-X}
\end{equation}
(and another: $\delta_+X$ with $+$ and $-$ interchanged).

PS proposed an algorithm for extending their analysis to higher orders
which can be stated as follows. Firstly, write down all possible
$(1,1)$ terms which according to PS simply amounts to keeping terms
whose net number of $\pm$-derivatives (terms in the denominator count
negatively) is $(1,1)$.  Secondly, discard all terms proportional to
the leading-order constraints $\del_\pm X\cdot\del_\pm X$ and their
derivatives. Finally, use integration by parts to relate equivalent
terms.

At this point one will have terms with and without `mixed
derivatives', terms sporting mixed derivatives being what we have
called \emph{irrelevant} in \cite{HDPM2}. The PS prescription then is
to discard all irrelevant terms and \emph{find transformation laws}
that leave the relevant terms in the action invariant.

Clearly, generalisation of the PS formalism requires not only finding
the right action to the desired order, but also determining the
appropriate transformation laws.  This is reminiscent of the early
days of supergravity theories, and this procedure becomes tedious and
unwieldy with increasing order. Not only does the procedure become
tedious, more importantly it does not lend itself to a systematic
method of construction and analysis at higher orders. It is the
purpose of the present paper to propose a simplified formalism, in
both a technical and conceptual sense. We propose to achieve this
through a formulation wherein the transformation laws are independent
of the particular action chosen. We start by demonstrating how this
can be done for the PS action itself.

Recall that the PS proposal for the leading correction was based on a
comparison with the Liouville action for subcritical strings
\begin{equation}
S_{Liou} = \frac{26-D}{48\pi}\int~d\tau^+~d\tau^-~~ \partial_+\phi\partial_-\phi
\end{equation}
They argued that in effective string theories the conformal factor
$e^\phi$ should be replaced by the component
$\partial_+X\cdot\partial_-X$ (in the conformal gauge) of the induced
metric on the worldsheet. They had also proposed replacing
$(26-D)/12$ by a parameter $\beta$ which was to be determined by
requiring the vanishing of the total central charge in all dimensions,
though they eventually found $\beta$ to be just the same as in the
Liouville theory \footnote{NDH~thanks Hikaru Kawai regarding why this has to be so.}. 
A direct application of this idea would have
suggested the total action
\begin{eqnarray}
\label{action2}
S_{2} &=& \frac{1}{4\pi} \int \td\tau^+ \td\tau^- \bigg\{
 \frac{1}{a^2} \del_+ X^\mu \del_- X_\mu \nn\\
&&\mbox{}+\beta 
\frac {\del_+(\del_+X\cdot\del_-X)\del_-(\del_+X\cdot\del_-X)}{(\del_+X\cdot\del_-X)^2}
\bigg\}
.\end{eqnarray}
It is easily shown that $S_{(2)}$ is invariant under
the transformations
\begin{equation}
\label{urtrans}
\delta^{0}_\pm X^\mu = \epsilon^\pm(\tau^\pm)\del_\pm X^\mu
.\end{equation}
For the purposes of the present discussion, we consider the `$-$'
alternative, without loss of generality.
More explicitly, if we write the second part of $S_2$, $S_2^{(2)}$, as
\begin{equation}
\label{defL2}
S^{(2)}_2 = \int \td\tau^+\td\tau^- {L}_2
\end{equation}
it is easy to show that
\begin{equation}
\label{psanom}
\delta^{0}_- L_2 = \del_-(\epsilon^- L_2) + \del_-^2\epsilon^- \del_+ L
\end{equation}
The first term is what one would have expected if $L_2$ had
transformed as a scalar density, and the second term is a departure
from this. We shall explain this important point later; for the moment
it suffices to note that the additional term can be rewritten as
\begin{equation}
\del_-^2\epsilon^- \del_+ L = \del_+(\del_-^2\epsilon^- L)
\end{equation}
ensuring the invariance of $S_2$ if we neglect integrals of total derivative terms. 

Polchinski and strominger \cite{PS} build their effective action while
discarding all total derivatives. This has generally been done in the
literature; treatment of total derivative terms in the action is a
subtle and important issue that in principle needs careful
scrutiny. In this paper, we shall nevertheless proceed with the
premise that such total derivative terms can be ignored.

The algebra of the PS transformations of \er{modtrans} is
\begin{equation}
[\delta^{PS}_-(\epsilon_1^-) , \delta^{PS}_-(\epsilon_2^-)] = \delta^{PS}_-(\epsilon_{12}^-) +\Pcm{O}(R^{-4})
,\end{equation}
where $\epsilon_{12}^- = \epsilon_1^-\del_-\epsilon_2^- - \epsilon_2^-\del_-\epsilon_1^-$. 
On the other hand, the algebra of the
transformations of \er{urtrans} is
\begin{equation}
[\delta^{0}_-(\epsilon_1^-) , \delta^{0}_-(\epsilon_2^-)] = \delta^{0}_-(\epsilon_{12}^-) 
.\end{equation}
Thus both generate the same group of symmetry transformations, namely
the conformal group. While the PS transformations realise this only
approximately, to $\Pcm{O}(R^{-4})$ which however is sufficient in
context as the PS action is defined to $\Pcm{O}(R^{-3})$, the
transformations \eqref{urtrans} leaving $S_{2}$ invariant do so \emph{exactly}.

It should be noted that field redefinitions do not change the algebra
of transformations, though the transformation laws are themselves
changed. This is indeed what is happening here and to understand this
note
\begin{equation}
\label{2ps}
S_{PS}-S_{2} = \frac{\beta}{4\pi}\int L^{-2}\partial_{+-}X\cdot\partial_-X
\partial_+L
\end{equation}
where $L=\partial_+X\cdot\partial_-X$. Thus the additional terms are
proportional to the EOM of the leading part of the action, and can be
removed through field redefinitions to appropriate order. A detailed
discussion of how the field redefinition corresponding to \er{2ps}
indeed connects \er{modtrans} and \er{urtrans}, as well as an
alternate description of effective string theories based on
the action $S_{2}$ can be found in \cite{HDPM2}.

\section{Simplified Formalism}
The above discussion points to a much simpler formulation of effective
string theories whereby the transformation law is always of the form
\eqref{urtrans}. Furthermore, $S_{2}$ provides an example of an
effective string theory which is in principle valid to all orders in
$1/R$, and would provide an important test case for understanding
higher-order corrections to the spectrum of Nambu-Goto theory.  Before
beginning construction, we briefly discuss here the merits of such a
formulation.

\subsection{Covariant Formalism}
Since the transformation is fixed and does not have to be fine-tuned
to a given action we now have the possibility of a systematic
covariant calculus for the construction of invariant actions. One is
always assured of obtaining invariant actions this way whereas a
generalisation of the PS algorithm based on the na\"ive $(1,1)$
counting there is no guarantee that to a given action one can always
find a suitable transformation law. In a given case one has either to
use trial and error or to identify a field redefinition, and even then
the results are valid only up to some prescibed order. On the other
hand each construction based on a covariant calculus will yield
actions valid to all orders in $1/R$. In the present paper, we
approach the construction of a covariant formulation in two ways,
explicitly given in sections \sref{covMet} and \sref{covWeyl}.

\subsection{Measures and Quantum Equivalences}
An important issue tied up with field redefinitions in Quantum Field
Theory is that of the quantum equivalence of theories related by
them. This has been addressed to some extent in \cite{HDPM2}. In the
path integral formulation, to which the canonical formulation should
be equivalent after all care has been exercised, this concerns the
invariant (under the symmetry transformations) measure to be adopted
as well as its transformation under field redefinitions. Both these
issues are naturally taken care of in a covariant formulation based on
\er{urtrans}. As the `na\"ive' measure $\prod_\sigma \td X(\sigma)$ is
invariant modulo irrelevant regularisation-dependent factors,
specifying the action specifies everything. This is a great
simplification. Covariance fixes the irrelevant terms also thereby
fixing a field definition also.

The last point also means that in the covariant formulation one cannot
simply drop the irrelevant terms as that would amount to changing the
field definition which generically would result in a change of
measure, as well as the transformation law which would necessitate
changing the covariant calculus itself. If, however, it can be shown
that the resulting field redefinition to a certain order does not
spoil quantum equivalence (i.e. the measure is left unchanged to the
relevant order), irrelevant terms can indeed be dropped.  However, the
resulting changes in the transformation law have to be taken into
account.

\section{Two paths to covariance}
\subsection{Symmetry Content}

It is the symmetry content of the theory, more precisely the symmetry
variations (transformation laws) that determine the covariant
calculus. Clearly, this is dictated by the physics of the system and
is not a matter of formalism. We are seeking a covariant formalism for
the symmetry variations of \er{urtrans}. Before doing so it is
worthwhile understanding why these should embody the symmetry content
of effective string theories.  Justifiably one could have taken the
view that this depends on the details of systems with effective string
behaviour. For example, it may have been so that only the `global'
version of \er{modtrans} as against the more restrictive `local'
version correctly captures the relevant symmetry. It just so happens
that for both the leading order effective action as well as for the PS
terms, the global invariance also implies the local invariance.
Clearly at high orders this will no longer be true. Then it will
become a matter of `phenomenology' to find out which will be a better
description. Nevertheless, we shall develop a covariant formulation
for the local invariance. Should phenomenology prefer the global
invariance as the true symmetry the rationale for such a covariant
formulation would be considerably weakened. It should be pointed out
that even then such a covariant formulation will be useful as a
framework for any systematic phenomenological analysis.

In what follows we shall actually seek something more general. We
shall seek the most general coordinate invariant version of the
transformation laws of \er{urtrans} and develop the
corresponding covariant calculus.
\subsection{Two paths}
\label{twopaths}
We have mentioned that we will construct our covariant formalism in
two alternate ways. The two distinct approaches are similar to what has been followed
in the case of fundamental string theories. 

The first, the Nambu-Goto method, is to
start with the action
\begin{equation}
\label{nambu} 
S_{\textup{NG}} = \int \sqrt{\det(\del_\alpha X\cdot\del_\beta X)}
,\end{equation}
invariant under \er{covtrans1}.
This approach is characterised by the absence of an intrinsic metric
on the worldsheet.  The composite operator, $\del_\alpha
X\cdot\del_\beta X$, also the induced metric on the worldsheet due to
the flat geometry of the target space, transforming exactly as the
metric, acts as a substitute metric in realising general coordinate
invariance.

The second, the Polyakov method, introduces an auxiliary metric field $h_{\alpha\beta}$.
The action equivalent to \er{nambu} is the Polyakov action
\begin{equation}
\label{polyact}
S=\int \td^2 \sigma \sqrt{h} h^{\alpha\beta} \del_\alpha X \cdot \del_\beta X
,\end{equation}
invariant under \er{covtrans1} and \er{covtrans2}. The metric field
$h_{\alpha\beta}$ is independent. The Polyakov action is also
general-coordinate invariant, although the real symmetry content is
reflected in the invariance of the action under the Weyl
transformations, $h_{\alpha\beta}\rightarrow \lambda(\sigma)
h_{\alpha\beta}$.

In effective string theories one will necessarily have to consider
higher derivative terms in the action and these may not in general be
Weyl invariant. This will require some additional technical structures
which are developed in \sref{covWeyl}. In fundamental string theories
one did not need these.

Although conceptually distinct, both these approaches lead to
identical content for the final effective string theory they are
designed to produce. This will be shown in detail in section
\sref{equiv}.

\subsection{Reparametrisation Invariance vs. Symmetry}
\label{concepts}
It can be seen from the above that reparametrisation invariance plays radically different
r\^oles in the two approaches. It is worthwhile understanding this important difference. 
Generically reparametrisation invariance is considered in situations with an intrinsic
metric for the space-time manifold. In such situations, any action can be made reparametrisation
invariant, and consequently the latter is devoid of physical content.
It is only the statement that specific choice of  coordinates is immaterial, and that it is
desirable to write the theory in a \emph{form} which reflects this. It is not a symmetry
of the physical system.\footnote{ 
Almost immediately after Einstein had formulated his General Theory of Relativity,
Kretschmann \cite{kretsch} had pointed this out but it is not widely appreciated
even at present.}
This is best illustrated by the following elementary example.
Consider a theory with a scalar field $\phi$ on a flat background. The action could look something like
\begin{equation}
\label{scalarex}
S=\int\td^d x \big( \del_\mu\phi \del_\mu\phi - m^2\phi^2 + \cdots + \phi^4 + \cdots \big)
.\end{equation}

Since it doesn't matter what coordinates we choose, 
it is desirable to represent the theory in
a way that under general coordinate transformations, the action is
invariant, in the sense that it \emph{has the same form in any
coordinate system}:
\begin{equation}
\label{scalarexgen}
S=\int\td^d x \sqrt{g} \big( g^{\mu\nu}\nabla_\mu\phi \nabla_\nu\phi - m^2\phi^2 + \cdots + \phi^4 + \cdots \big)
.\end{equation}

It should be noted that this does not \emph{change} the theory at all. 
The crucial point is that this applies to \emph{any} theory. 
Any theory can thus
be written in such diffeomorphism-invariant form, so that a
particular choice of coordinates can be postponed or avoided.

Turning our attention to symmetries, the situation is conceptually
different. Symmetries, unlike diffeomorphisms in the above context,
restrict the physical content of the theory and not any theory can be
made invariant under the symmetry transformations.  A trivial example
in the context of the above mentioned scalar field theory is the
symmetry under $\phi\rightarrow -\phi$. This restricts the form of the
action to have only even powers of $\phi$ and not all actions possess
this feature irrespective of the choice of coordinates.

While any \emph{theory} must be diffeomorphism invariant, and
therefore can be written down in a covariant way which reflects this,
only certain theories have a particular symmetry. There is no way to
take an arbitrary theory and somehow \emph{make} it symmetric.

\subsection{Conformal Symmetry from Reparametrisation Invariance}
\label{demon}

As will be seen later, in the first approach conformal symmetry emerges as residual
invariance of the conformal gauge choice \er{cgauge1}. Since in this approach this
symmetry arises from the underlying reparametrisation invariance, which has been argued
above to be generically void of physical content, it is important to understand the 
precise connection between this emergent conformal symmetry and reparametrisation
invariance.

Does the group of reparametrisations contain the group of conformal
transformations?  Strictly, it does indeed; A mapping (assumed
invertible, differentiable, etc..)  from $x$ coordinates to $x'$ is a
general coordinate transformation, and of course the mappings which
correspond to conformal transformations are of the same kind.  It is
crucial to realise that reparametrisation invariance does not always
result in conformal symmetry upon choosing the conformal gauge.

This is best exemplified again by the scalar field example of
\er{scalarexgen} in two dimensions where coordinates can generically
(at least locally) be chosen so that the intrinsic metric is of the
form
\begin{equation}
\label{confgmet}
g_{\alpha\beta} = \left[ \begin{array}{cc} 0 & \varphi \\ \varphi & 0 \end{array}\right]
\end{equation}
in coordinates $\sigma^\pm$.  This does not use up all available
freedom, and residual coordinate transformations which preserve this
form are easily seen to comprise the conformal group. The action of
\er{scalarexgen} is indeed invariant under the action of these
transformations; yet the physical content of the theory is exactly
that of \er{scalarex}. What is more, \emph{any} scalar field theory
can be made to have this invariance, and it therefore does not
represent any physical symmetry.

In the second approach what does represent a symmetry is the
invariance under Weyl-scaling.  In the scalar field example also one
sees that not all actions possess this invariance in keeping with what
a symmetry is.

In our first approach, in which we do not treat the metric as an
independent field, we do not make any assumption of Weyl symmetry.
Nevertheless there is a symmetry in this case and that is traceable
entirely to reparametrisation invariance. The r\^ole of an intrinsic
metric is instead played by suitable composite fields constructed out
of the physical fields.  The only degrees of freedom in the theory are
taken to be the $X$ scalar fields.  Since now not every action can be
reparametrisation invariant, reparametrisation invariance in this case
becomes a physical symmetry.

Going to the equivalent of \eqref{confgmet} by a coordinate choice,
where $g_{\alpha\beta}$ is now a composite field transforming like a
metric, and choosing
\[
\varphi \equiv \del_+ X \del_- X
\]
one realises the conformal gauge of the first formalism with conformal
invariance as the residual symmetry.

Thus in both approaches conformal invariance emerges as the residual
invariance of the conformal gauge; but in the first case it emerges as
a true physical symmetry, while in the second approach it is like a
generic reparametrisation invariance but not a symmetry. It is the
underlying Weyl-scaling invariance that finally results in the
conformal invariance being elevated to a symmetry in the precise sense
that not all actions are invariant.  It should also be emphasised that
in other gauges, like for example the transverse gauge $X^0=\tau,
X^1=\sigma$ there will be nothing like conformal symmetry in either of
the two approaches. In that sense, this is true of fundamental string
theory also, there is nothing intrinsic to conformal symmetry per se;
what is important is the symmetry content of the gauge-unfixed theory.

\subsection{The Denominator Principle}
\label{denomprin}

What is being developed in this work is for effective string theories
as opposed to fundamental string theories. The allowed actions for
effective string theories can sometimes become singular for certain
string configurations but for long strings fluctuating about a
classical background such action terms should be sensible. However
even this requirement should preclude terms in effective string
actions whose denominators can become singular for some flucuation of
the effective string. This becomes an important guiding principle for
effective string theories. In particular, it needs to be evoked while
restricting substitute metrics in section \sref{covMet}, the Weyl
connection in section \sref{covWeyl} as well as restricting the
Weyl-weight compensators in section \sref{covacts2}.
\section{Covariant Calculus I: non-intrinsic metric}
\label{covMet}
In this section we make one of our proposals for a covariant
formulation. To attain final covariance under conformal
transformations, we shall use initially covariance under worldsheet
general coordinate transformations only.  \emph{A priori}, a metric
field is needed for any covariant formulation. In the spirit of PS we
shall not introduce any intrinsic metric on the worldsheet. It
suffices to have an object that transforms the same way as a metric
under general coordinate transformations. One natural choice for such
a \emph{metric substitute} is the {\it induced metric} on the worldsheet

\begin{equation}
\label{induced}
g_{\alpha\beta} = \del_\alpha X\cdot\del_\beta X
.\end{equation}

Strictly speaking, any quantity built out of the basic variables
$X^\mu$ with the correct $2-d$ tensor structure is also a \emph{bona
  fide} candidate. In fact, any such object would lead to a
formulation in which covariance is manifest, and effective actions
could be constructed.  The choice of \er{induced} is in a sense the
\emph{simplest} one can make and it is also the choice that PS made.

Finally, the quantity we choose here will later appear, in gauge-fixed
form, in various denominators. As we require the effective theory to
be valid on any fluctuation, \er{induced} is the simplest choice, just
as $L$ was for denominators in the initial PS formulation and
subsequent treatments \cite{Drum,HDPM2}. These choices are also
consistent with the Denominator Principle enunciated above.  All other
choices, upon resorting to perturbation in $R^{-1}$, are essentially
equivalent to this.

Once the metric substitute is chosen, the rest of the construction is
along standard lines of Riemannian Geometry. Various covariant
derivatives $D_{\alpha\beta\gamma..}X$ can be written, and invariants
made out of the $g$ and these objects. In addition, tensors containing
only the derivatives of $g$ can only enter through the Riemann
curvature tensor $R_{\alpha\beta\gamma\delta}$ and its covariant
derivatives. Since in two dimensions
\begin{equation}
\label{rtensor}
R_{\alpha\beta\gamma\delta} = (g_{\alpha \gamma}g_{\beta\delta} - g_{\beta\gamma}g_{\alpha\delta})\frac{R}{2}
,\end{equation}
where $R$ is the Ricci scalar, one need consider $R$ and its covariant
derivatives only. This vastly simplifies the construction of actions.

\subsection{Some Manifestly Covariant Actions - I}
\label{covacts1}
In this section we provide a few examples of manifestly covariant
action terms, more specifically, terms that transform as scalar
densities. A systematic procedure for construction of such terms to
any desired order in $1/R$ will be given later, in section
\sref{highercovterms}. One could begin with
\begin{equation}
I_{\textup{cov}} = \sqrt{g}D_{\alpha_1\beta_1..}X^{\mu_1}D_{\alpha_2\beta_2..}X^{\mu_2}\cdot A^{\alpha_1\beta_1\cdots\alpha_2\beta_2\cdots}B_{\mu_1\mu_2\cdots}
\end{equation}
where $A^{\alpha_1\beta_1\cdots\alpha_2\beta_2\cdots}$ is composed of
suitable factors of Levi-Civita and metric tensors on the
two-dimensional worldsheet and $B_{\mu_1\mu_2\cdot}$ made up of
$\eta_{\mu\nu}$ and Levi-Civita tensors in target space. In the
conformal gauge this construction can be done even more simply by
stringing together a number of covariant derivatives so that there are
equal net numbers of $(+,-)$ indices, and finally use sufficient
inverse powers of $L$ to make the expression $(1,1)$.

Now we illustrate these methods by covariantising some terms proposed
by Drummond. The PS term itself is at leading order $R^{-2}$, and
Drummond \cite{Drum} found four possibilities for the next relevant
order-$R^{-6}$ part of the action. These are
\begin{eqnarray}
\label{dterms1}
M_1 & = & \frac{1}{L^3} \del_+^2X\cdot\del_+^2X~\del_-^2X\cdot\del_-^2 X ,\\
\label{dterms2}
M_2 & = & \frac{1}{L^3} \del_+^2X\cdot\del_-^2X~\del_+^2X\cdot\del_-^2 X ,\\
\label{dterms3}
M_3 &=& \frac{1}{L^4} \del_-^2 X \cdot \del_+^2 X \del_- X \cdot \del_+^2 X \del_-^2 X \cdot \del_+ X, \\
\label{dterms4}
M_4 &=& \frac{1}{L^5} (\del_- X \cdot \del_+^2 X)^2 (\del_-^2 X \cdot \del_+ X)^2
.\end{eqnarray}

Considering the first two terms, we can expect these to be contained
in the covariant forms
\begin{equation}
\label{cov1ex1}
{\cal M}_1  =  \sqrt{g}D_{\alpha_1\beta_1}X\cdot D_{\alpha_2\beta_2}X~D^{\alpha_1\beta_1}X\cdot D^{\alpha_2\beta_2}X 
\end{equation}
\begin{equation}
\label{cov1ex2}
{\cal M}_2  =  \sqrt{g}D_{\alpha_1\beta_1}X\cdot D^{\alpha_1\beta_1}X~D_{\alpha_2\beta_2}X\cdot D^{\alpha_2\beta_2}X
\end{equation}
\subsection{Conformal Gauge and Conformal Transformations in Calculus-I}
\label{conf1}
The PS formulation specifically hinged on the use of the conformal transformations
\begin{equation}
\label{conftr}
\tau^\pm~\rightarrow \tau^\pm + \epsilon^\pm; \qquad \quad \del_\pm~\epsilon^\mp = 0
\end{equation}
In the context of general coordinate invariance, these transformations
arise as the residual transformations maintaining the conformal gauge
\begin{equation}
\label{cgauge1}
g_{++} = g_{--} =0
\end{equation}
In this gauge $g_{+-}=g_{-+}=L$ transforms as a true $(1,1)$-tensor
under the conformal transformations. Importantly,
$g^{+-}=g^{-+}=L^{-1}$ transforms as a $(-1,-1)$ tensor.  It is
straightforward to work out the non-vanishing components of the
Christoffel connection as well as the Riemann curvature tensor:
\begin{equation}
\label{Gamma1}
{\Gamma^{(1)}}^+_{++} = \del_+ \ln L;~~~~{\Gamma^{(1)}}^-_{--} = \del_- \ln L
\end{equation}
\begin{eqnarray}
R^+_{+-+} &=& -R^+_{++-} = \del_+\del_- \ln L\nn\\
R^-_{-+-} &=& -R^-_{--+} = \del_+\del_- \ln L
\end{eqnarray}
All the remaining components are zero.  The resulting scalar curvature
is
\begin{equation}
\label{ricciscalar}
R = -2 \frac{\del_+\del_- \ln L}{L};~~~~~~\sqrt{g} R = -2\del_+\del_- \ln L
\end{equation}
We next give explicit expressions for some covariant derivatives:
\begin{eqnarray}
\label{explicit}
D_\pm~X^\mu &=& \del_\pm~X^\mu\nn\\
D_{++}X^\mu &=& \del_{++}X^\mu - \del_+\ln L\del_+ X^\mu\nn\\
D_{--}X^\mu& =& \del_{--}X^\mu - \del_-\ln L\del_-X^\mu\nn\\
D_{+-} X^\mu &=& D_{-+}X^\mu = \del_{+-}X^\mu\nn\\
D_{++-} X^\mu &=& D_{+-+} X^\mu = \del_{++-} X^\mu -\del_+\ln L~\del_{+-} X^\mu\nn\\
D_{-++} X^\mu &=& \del_{-++} X^\mu - \del_-(\del_+\ln L~\del_{+} X^\mu)  
\end{eqnarray}
The last two of these equations show that i) just the number of $\pm$ indices does not fully characterise
a tensor; their order is important. ii) not all tensors with mixed indices are proportional to leading 
order EOM. The latter will alter the rules for constructing general actions in comparison to what was
discussed in \cite{HDPM2,Drum2}. However the last but one equation displays mixed-indices tensors
that are indeed proportional to leading order EOM. This is a consequence of the following two important
relations:

If $T_{\mu_1\dots\mu_n}$ is a tensor with $m_\pm$ indices of type $\pm$,
\begin{eqnarray}
D_{+}T_{\mu_1\dots\mu_n} &=& \del_+~T_{\mu_1\dots\mu_n} - m_+\del_+\ln L  T_{\mu_1\dots\mu_n}\nn\\
D_{-}T_{\mu_1\dots\mu_n} &=& \del_-~T_{\mu_1\dots\mu_n} - m_-\del_-\ln L  T_{\mu_1\dots\mu_n}\nn\\
\end{eqnarray}
Hence covariant derivatives of tensors which are a combination of leading order EOM and its
derivatives are also combinations of leading order EOM and its derivatives.

Another important property is that
$D_{\pm\pm}X\cdot D_\pm X$ are linear combinations of leading order constraints
$\del_\pm X\cdot\del_\pm X$ and their derivatives. That is,
\begin{equation}
\label{pmpmconst}
D_{\pm\pm}X\cdot D_\pm X = \frac{1}{2}\del_\pm(\del_\pm X\cdot\del_\pm X) - \del_\pm\ln L (\del_\pm X\cdot
\del_\pm X)
.\end{equation}
This too follows trivially from the second and third eqns of \er{explicit}.
\section{Covariant Calculus II: Intrinsic Metric and Weyl Symmetry}
\label{covWeyl}

In this section, we develop the covariant calculus based on the
Polyakov approach which is both general coordinate invariant and
Weyl-invariant.

We constuct covariant derivatives with respect not only to the
diffeomorphisms, but also to the Weyl-scaling symmetry, and use these objects
to construct covariant terms. Although this approach is
quite different from the non-intrinsic metric approach of section
\sref{covMet}, we will show in the end that the two approaches give
identical results.

\subsection{conformal symmetry}
\label{review}
Beginning with the Polyakov action
\begin{equation}
\label{polyact2}
S=\int \td^2 \sigma \sqrt{h} h^{\alpha\beta} \del_\alpha X \cdot \del_\beta X
\end{equation}
since $X^\mu$ is a worldsheet scalar, this construction ensures
two-dimensional worldsheet reparametrisation invariance. The
infinitesimal such transformation generated by 
$\sigma \rightarrow \sigma' = \sigma - \epsilon (\sigma)$ 
is given by
\begin{equation}
\label{covtrans1}
\delta_\epsilon X^\alpha = \epsilon^\gamma \del_\gamma X^\alpha
.\end{equation}
\begin{equation}
\label{covtrans2}
\delta_\epsilon h^{\alpha \beta} = 
\epsilon^\gamma\del_\gamma h^{\alpha\beta}
-\del_\gamma\epsilon^\alpha h^{\gamma\beta}
-\del_\gamma\epsilon^\beta h^{\alpha\gamma} 
.\end{equation}
The  important symmetry of \eqref{polyact2} is of course the 
local Weyl Scaling, which only affects the metric,
\begin{equation}
\label{finWeyl}
h_{\alpha\beta} \rightarrow h'_{\alpha\beta} = \omega(\sigma) h_{\alpha\beta}
\end{equation}
whose infinitesimal version with $\omega(\sigma)=1+\lambda(\sigma)$ reads 
\begin{equation}
\label{inflWeyl}
\delta_\lambda h_{\alpha\beta} = \lambda h_{\alpha\beta}
.\end{equation}

A combination of the
reparametrisation and Weyl symmetries is used to bring the worldsheet metric
to the form $h_{\alpha\beta}=\eta_{\alpha\beta}$, called conformal gauge. 

This choice of $h_{\alpha\beta}$ does not fix the coordinates and the
freedom of Weyl scaling completely; A combined Weyl scaling and
coordinate transformation such that
\begin{equation}
\label{wctcomb}
\lambda h_{\alpha\beta} = \del_\beta\epsilon_\alpha + \del_\alpha\epsilon_\beta 
\end{equation}
preserves $h_{\alpha\beta}=\eta_{\alpha\beta}$.
This residual symmetry is worldsheet conformal symmetry.
Defining coordinates $\tau^{\pm} = \tau \pm \sigma$, the remaining 
infinitesimal symmetries are parametrised by arbitrary functions
\begin{equation}
\epsilon^+(\tau^+) , \qquad \textup{and} \qquad \epsilon^-(\tau^-) 
.\end{equation}
This is of course just the symmetry of \er{urtrans}.

\subsection{Generalised Covariant Derivatives }
\label{covf}

What we need are quantities that transform covariantly under both general
coordinate transformations as well as local Weyl scalings. Hence we need
tensors with definite Weyl-scaling dimensions. A tensor $\phi$ of Weyl-scaling
dimension $j$ transforms under Weyl-scalings as
\begin{equation}
\label{finWeyl2}
\phi \rightarrow \phi '= \omega(\sigma)^j \phi 
\end{equation}
The Weyl-weight of $h_{\alpha\beta}$ is $1$ according to \er{finWeyl}
(this is a matter of convention without any loss of generality).  To
see the issues involved, consider a worldsheet vector $V_\beta$ with
Weyl-weight $j_V$; its covariant derivative with respect to
reparametrisations is
\begin{equation}
\label{covD1}
\nabla_\alpha V_\beta = \del_\alpha V_\beta - \Gamma_{\alpha\beta}^\gamma V_\gamma
\end{equation}
with the connection $\Gamma_{\alpha\beta}^\gamma$ given by the standard Christoffel symbol
\begin{equation}
\label{connection}
\Gamma_{\alpha\beta}^\gamma =  -\frac12 h^{\gamma\delta} \big(
\del_\delta h_{\alpha\beta}-\del_\alpha h_{\beta\delta}-\del_\beta h_{\alpha\delta}
\big)
.\end{equation}

Clearly under Weyl-scalings of $V_\alpha$ the covariant derivative of
\er{covD1} does not scale in any simple way. In this particular
example, there are two sources for this; the occurrence of the
derivative of $V$ on the one hand, and the occurrence of the
derivatives of $h_{\alpha\beta}$ on the other. Ordinary derivatives of
a tensor $\phi$ with definite Weyl-weight $j$ do not simply scale when
$\phi$ is locally scaled.

This motivates the definition of a new \emph{Weyl-covariant derivative}; for a tensor field
$\phi$ of 
Weyl-scaling dimension $j$, we set
\begin{equation}
\label{weylD}
\Delta_\alpha \phi \equiv \del_\alpha \phi - j \chi_\alpha \phi
\end{equation}
Restricting to the case when $\phi$ is a scalar, one sees that
$\chi_\alpha$ must transform as a worldsheet vector under
reparametrisations. The Weyl-covariant derivative of a field with
Weyl-scaling dimension $j$ should again be a field with the same
Weyl-scaling dimension $j$:
\begin{equation}
\label{wcdscalar}
(\Delta_\alpha \phi) '= \omega^j \Delta_\alpha \phi
\end{equation}
under the transformation \eqref{finWeyl}. This requires the following
inhomogeneous transformation of $\chi_\alpha$ under Weyl-scaling
\begin{equation}
\label{chitrans}
\chi'_\alpha = \chi_\alpha + \del_\alpha \ln\omega
\end{equation}

This immediately leads to the following generalisation of the Christoffel
symbol that is appropriate for the present context:
\begin{eqnarray}
\label{Gconnection}
G_{\alpha\beta}^\gamma &=& \frac12 h^{\gamma\delta} 
\big(
 \Delta_\alpha h_{\beta\delta}
 +\Delta_\beta h_{\alpha\delta}
-\Delta_\delta h_{\alpha\beta}
\big)\nn\\
&\equiv& \Gamma_{\alpha\beta}^\gamma + W_{\alpha\beta}^\gamma
\end{eqnarray}
where
\begin{equation}
\label{Wtensor}
W_{\alpha\beta}^\gamma 
= \frac12(h_{\alpha\beta}\chi^\gamma - \delta_\alpha^\gamma \chi_\beta - \delta_\beta^\gamma \chi_\alpha  ) 
.\end{equation}

From \er{Gconnection} it is easy to see that $G_{\alpha\beta}^\gamma$
is invariant under Weyl-scalings (it has Weyl-weight $0$) while
neither $\Gamma$ nor $W$ has well-defined Weyl-weight. Since
$W_{\alpha\beta}^\gamma$ transforms as a proper tensor under
reparametrisations, it follows that $G_{\alpha\beta}^\gamma$ also
transforms as a proper connection.  Putting these observations
together, we define the Weyl-reparametrisation covariant derivative
$\Pcm{D}_\alpha$ of a rank-$n$ worldsheet tensor
$T_{\beta_1\dots\beta_n}$ of Weyl-scaling dimension $j$ by
\begin{eqnarray}
\label{totcovder}
\Pcm{D}_\alpha T_{\beta_1\dots\beta_n} &\equiv&
\Delta_\alpha T_{\beta_1\dots\beta_n} \nn\\
&-& G_{\alpha\beta_1}^\gamma T_{\gamma\beta_2\dots\beta_n} \nn\\
&-& \cdots \nn\\
&-& G_{\alpha\beta_n}^\gamma T_{\beta_1\dots\beta_{n-1}\gamma} 
\end{eqnarray}
It is useful to rewrite this in the suggestive form
\begin{eqnarray}
\label{totcovder2}
\Pcm{D}_\alpha T_{\beta_1\dots\beta_n} &\equiv&
D_\alpha T_{\beta_1\dots\beta_n} -j\chi_\alpha T_{\beta_1\dots\beta_n}\nn\\
&-& W_{\alpha\beta_1}^\gamma T_{\gamma\beta_2\dots\beta_n} \nn\\
&-& \cdots \nn\\
&-& W_{\alpha\beta_n}^\gamma T_{\beta_1\dots\beta_{n-1}\gamma} 
\end{eqnarray}
In \er{totcovder} every term has the same Weyl-weight as the tensor
$T$ and consequently so does $\Pcm{D} T$, but none of these terms
transforms as a tensor under reparametrisations. On the other hand in
\er{totcovder2} every term transforms as a tensor under
reparametrisations while none of them has a definite
Weyl-weight. Together equations \eqref{totcovder} and
\eqref{totcovder2} imply that $\Pcm{D} T$ is covariant under both
Weyl-scalings and reparametrisations.

The various covariant derivatives obey a
Leibniz rule, just as $\del_\alpha$ does:
\begin{eqnarray}
\label{leibnitz}
\nabla_\alpha ( T_1 T_2 ) &=& \nabla_\alpha T_1 T_2 + T_1 \nabla_\alpha T_2 ,\\
\Delta_\alpha ( T_1 T_2 ) &=& \Delta_\alpha T_1 T_2 + T_1 \Delta_\alpha T_2 ,\\
\Pcm{D}_\alpha ( T_1 T_2 ) &=& \Pcm{D}_\alpha T_1 T_2 + T_1 \Pcm{D}_\alpha T_2
.\end{eqnarray}
where $T_1$ and $T_2$ are tensors, each with definite Weyl dimension, but not necessarily
of the same rank.

$\Pcm{D}$ sports the important property
\begin{equation}
\label{dhpzero}
\Pcm{D}_\alpha h_{\gamma\delta} = 0 
.\end{equation}

\subsection{Weyl Connection}
\label{weylconnect}
All the features discussed above hold for any choice of $\chi_\alpha$ as long it responds to
Weyl-scalings according to \er{chitrans}. In fact, according to that equation, a connection 
of the form
\begin{equation}
\label{connectform}
\chi_\alpha = \frac1{W_\Phi} \del_\alpha \log \Phi
\end{equation}
where $\Phi$ is any worldsheet \emph{scalar} of Weyl-scaling dimension
$W_\Phi$, would be acceptable. It follows that
\begin{equation}
\Pcm{D}_\alpha \Phi = 0
.\end{equation}
We shall choose $\Phi$ to be constructed from $h$ and derivatives of
$X$.

We are still free to choose a form for $\Phi$.  We are not constrained to
use only one form for $\Phi$; anything will do so long as it is a
scalar with non-zero Weyl-dimension, and also that it is conformity
with the Denominator Principle of \sref{denomprin}.  This constrains
$\Phi$ to be of the form
\begin{equation}
\Phi = \Pcm{L} + \textup{higher order in $1/R$}.
\end{equation}
where
\begin{equation}
\Pcm{L} \equiv h^{\alpha\beta} \Pcm{D}_\alpha X  \Pcm{D}_\beta X \equiv X_{;\alpha} X_{;\alpha}
\end{equation}

By arguments identical to the ones that led to \er{induced} as the simplest
choice for the metric substitute in the first approach, we conclude that the
simplest choice for $\Phi$ is
\begin{equation}
\label{Phichoice}
\Phi = \Pcm{L} \qquad W_\Phi = -1
\end{equation}
In section \sref{equiv} we shall see that there is indeed an intimate
connection between these two choices.

\subsection{Manifestly Covariant Action Terms - II}
\label{covacts2}
After constructing all the Weyl-reparametrisation covariant
derivatives $\Pcm{D}_{\alpha\beta\dots}~X^\mu$ with Weyl-weight $0$,
the Weyl-reparametrisation covariant generalised Riemann tensor
\begin{equation}
\label{griemann}
\Pcm{R}_{\beta\gamma\delta}^\alpha\equiv \Delta_\gamma G_{\beta\delta}^\alpha
-\Delta_\delta G_{\beta\gamma}^\alpha +G_{\gamma\eta}^\alpha G_{\delta\beta}^\eta
-G_{\delta\eta}^\alpha G_{\gamma\beta}^\eta
\end{equation}
and its Weyl-reparametrisation covariant derivatives, all of
Weyl-weight $0$, one can construct action integrands which are scalar
densities under reparametrisation and invariant under
Weyl-scalings. We shall do this as a two-step process to highlight
important differences from the corresponding construction in section
\sref{covacts1}; first we shall construct scalar densities under
reparametrisation and use Weyl-scaling covariance to eventually obtain
our quantities of interest. The first step is very similar to what was
done in section \sref{covacts1}. Let us illustrate this by working out
the analog of \er{cov1ex2} of section \sref{covMet};
\begin{equation}
\label{cov2ex0}
\bar{\Pcm{N}}_1 = \sqrt{h}\{ \Pcm{D}_{\alpha_1\beta_1}X\cdot\Pcm{D}_{\alpha_2\beta_2}X 
h^{\alpha_1\alpha_2}h^{\beta_1\beta_2}\}^2
\end{equation} 
The Weyl-weight of $\bar{\Pcm{N}}_1$ is $-3$ and that brings us to the second step; in order
to get a term of Weyl-weight $0$ one has to multiply by something with Weyl-weight $3$.
Clearly there are many ways of doing so. We call these {\it Weyl-weight Compensators}.
We now show that if the `total divergence' property of covariant derivatives is to be
extended to the Weyl-reparametrisation covariant derivatives, these compensators have
to be appropriate powers of $\Phi$ of \er{connectform}. 

Consider a contravariant vector $V^\alpha$ of Weyl-weight $J$. Its Weyl-reparametrisation
covariant derivative is given by
\begin{equation}
\Pcm{D}_\alpha V^{\beta} = \nabla_\alpha V^\beta - J \chi_\alpha V^\beta + W_{\alpha\gamma}^\beta V^\gamma
.\end{equation}
Hence
\begin{equation}
\Pcm{D}_\alpha V^{\alpha} = \nabla_\alpha V^\alpha - J \chi_\alpha V^\alpha + W_{\alpha\gamma}^\alpha V^\gamma
.\end{equation}
On recalling $\nabla_\alpha V^\alpha = \frac{1}{\sqrt{h}}\del_\alpha (\sqrt{h}V^\alpha)$
and $W_{\alpha\gamma}^\gamma = -\chi_\gamma$, one gets
\begin{equation}
\label{totder2}
\Pcm{D}_\alpha V^\alpha = \frac{\Phi^{\frac{J+1}{W_\Phi}}}{\sqrt{h}}\del_\alpha 
(\sqrt{h}\Phi^{-{\frac{J+1}{W_\Phi}}}V^\alpha)
\end{equation}
Thus in order to convert the scalar density $\sqrt{h}\Pcm{D}_\alpha
V^\alpha$ of Weyl-weight $J+1$ into a scalar density with Weyl-weight
$0$ so that the total divergence property is maintained, it has to be
multiplied only by $\Phi^{-(J+1)/W_\Phi}$ and not by just any
expression with Weyl-weight $-(J+1)$. In other words, the Weyl-weight
Compensators have to be appropriate powers of $\Phi$.  With the
specific choice of \er{Phichoice} these compensators are powers of
$\Pcm{L}$. Thus the final desired expression for our example is
\begin{equation}
\label{cov2ex}
{\Pcm{N}}_1 = \sqrt{h}{\Pcm{L}}^{-3}\{ \Pcm{D}_{\alpha_1\beta_1}X\cdot\Pcm{D}_{\alpha_2\beta_2}X 
h^{\alpha_1\alpha_2}h^{\beta_1\beta_2}\}^2
\end{equation} 
We will show later that \er{cov2ex} and eqn \er{cov1ex2} are the same.

\subsection{Conformal Gauge and Conformal Transformations in Calculus-II }
\label{conf2}

As explained in detail above, we begin with both Weyl and coordinate
invariance and intend to fix both to end up with something written in
``$+/-$'' notation. The resultant actions will be invariant under the
conformal transformation \eqref{urtrans}.

A choice of coordinates $\sigma^\pm$ and Weyl scaling \eqref{finWeyl} is made to set
\begin{equation}
\label{cgauge2}
h_{+-} = h_{-+} = 2 , \qquad  h^{+-} = h^{-+} = 1/2 
.\end{equation}

We write gauge-fixed quantities using a `check' and covariant quantities in script letters. 
For example,
\begin{equation}
\Pcm{L} \equiv h^{\alpha\beta} \Pcm{D}_\alpha X  \Pcm{D}_\beta X \equiv X_{;\alpha} X_{;\alpha}
\quad\rightarrow\quad  \check{\Pcm{L}} \sim L \equiv \del_+ X \del_- X
.\end{equation}
here we write the covariant $\Pcm{D}$ derivative with a ``$;$'' to save space, 
and also assume that repeated indices are summed using the metric. 

In this gauge ${\Gamma^{(2)}}^\gamma_{\alpha\beta}=0$ and the W-tensor is given by
\begin{equation}
\check{W}_{\alpha\beta}^\gamma = -\frac{1}{2L} 
\big( h_{\alpha\beta} \del^\gamma - \delta_\alpha^\gamma\del_\beta - \delta_\beta^\gamma\del_\alpha \big) L
.\end{equation}

\subsection{$+$ and $-$ skeletal forms}
\label{pmsf}

In this section we explore some of the consequences of this gauge fixing. Suppose
$T_{\alpha\beta\dots}$ is a gauge-fixed tensor of Weyl-dimension $j$.
\begin{equation}
\Pcm{D}_+ T_{\dots}^{(j)} = \del_+ T_{\dots} - j \chi_+ T_{\dots} - t_+ W_{++}^+ T_{\dots}
\end{equation}
where we have used that $W_{++}^- = 0$ and $W_{+-}^\pm=0$, and $t_+$ is the number of $+$ indices on $T$.
Evaluating the gauge-fixed $W$-connection, the only components which do not vanish are
\begin{equation}
\label{Wfix}
W_{++}^+=- \chi_+ = \del_+\ln L \qquad W_{--}^-=- \chi_- = \del_-\ln L
\end{equation}
and thus
\begin{equation}
\Pcm{D}_+ T_{\dots}^{(j)} =  \del_+ T_{\dots} - j \chi_+ T_{\dots} + t_+ \chi_+ T_{\dots}
.\end{equation}
Similarly for $+ \leftrightarrow -$,
\begin{equation}
\Pcm{D}_- T_{\dots}^{(j)} =  \del_- T_{\dots} - j \chi_- T_{\dots} + t_- \chi_- T_{\dots}
.\end{equation}
Evidently,
\begin{eqnarray}
\label{dcomm}
[\Pcm{D}_+,\Pcm{D}_-] T_{\dots}^{(j)} &=& (t_- -t_+) (\del_+\chi_-) T_{\dots}^{(j)} \\
&=& (t_- - t_+) (\del_+\del_-\log \Phi) T_{\dots}^{(j)}
,\end{eqnarray}
in fact consistent with our earlier calculations, despite the difference in formalism.

We now show that the Weyl-reparametrisation covariant derivatives $\Pcm{D}_{\alpha\beta\dots}~X^\mu$ are
identical to the covariant derivatives $D_{\alpha\beta\dots}~X^\mu$ of section \sref{covMet}. We show this
recursively by first proving that covariant derivatives of zero Weyl-weight tensors are the same in
both methods. Consider such zero weight tensors $T_{\beta_1\dots\beta_n}$. Then
\begin{eqnarray}
\label{identity}
\Pcm{D}_\alpha~T_{\beta_1\dots\beta_n} &=& 
\del_\alpha T_{\beta_1\dots\beta_n}-G^\gamma_{\alpha\beta_1}T_{\gamma\beta_2\dots}-\dots\nn\\
&=&
\del_\alpha T_{\beta_1\dots\beta_n}-{\check W}^\gamma_{\alpha\beta_1}T_{\gamma\beta_2\dots}-\dots\nn\\
&=&
\del_\alpha T_{\beta_1\dots\beta_n}-{\Gamma^{(1)}}^\gamma_{\alpha\beta_1}T_{\gamma\beta_2\dots}-\dots\nn\\
&=& D_\alpha~T_{\beta_1\dots\beta_n}
.\end{eqnarray}
We have used the important fact that the components of gauge fixed
W-tensor given by \er{Wfix} are identical to those of the Christoffel
connection of section \sref{covMet} given in \er{Gamma1} and that the
Christoffel symbols $\Gamma^{(2)}$ of covariant calculus-II in its
conformal gauge are all $0$. In other words, the
$G^\gamma_{\alpha\beta}$ is the same in the two conformal gauges. This
is easily understood as the metric choices of \er{cgauge1} and
\er{cgauge2} are related by the Weyl-scaling factor $L$, and the
tensor $G^\gamma_{\alpha\beta}$ is itself of Weyl-weight $0$.

An immediate and important corollary is that all the components of generalised Riemann tensor $\Pcm{R}^\alpha
_{\beta\gamma\delta}$ of the second approach are identical to the standard Riemann tensor $R^\alpha_{\beta\gamma
\delta}$ of section \sref{covMet}. In the light of \er{identity} all the covariant derivatives of the
Riemann tensors are also the same in the two approaches. This means that the tensor ingredients of the two
approaches in their conformal gauges are the same. However, what are different are the metric tensors needed
to construct scalar densities, and the compensators in the second approach. We shall however show in section
\sref{equiv} that even these conspire to match perfectly. It is shown in that section that this is not just
an accident of the choices made in \er{induced} and \er{Phichoice} but is a more general feature. 
It is also shown in that section that the said equivalence continues to hold even when action terms are
constructed using tensors of nonzero Weyl-weight.
\section{Equivalence of the two conformal gauge formalisms}
\label{equiv}
We shall now show that the two conformal gauge formalisms are equivalent and that this equivalence is
more general than the explicit choices made in \er{induced} and \er{Phichoice}. We illustrate this
equivalence by again considering the covariant actions of \er{cov1ex2} and \er{cov2ex}. We
have already shown that all the covariant derivatives of $X^\mu$ are all the same. Evaluating \er{cov2ex}
in the conformal gauge of \er{cgauge2} one gets
\begin{equation}
\label{2exconf}
{\check{\Pcm{N}}_1 = 
\frac{1}{2} L^{-3}(D_{++}X\cdot D_{--}X + D_{+-}X\cdot D_{+-}X})^2
\end{equation}
On the other hand evaluating \er{cov1ex2} in the conformal gauge \er{cgauge1} one gets
\begin{equation}
\label{1ex2conf}
{\check{\Pcm{M}}_2 = 4 L^{-3}
(D_{++}X\cdot D_{--}X + D_{+-}X\cdot D_{+-}X})^2
\end{equation}
Thus the two terms are equal modulo an irrelevant constant. 

Instead of the special choices \er{induced} and \er{Phichoice} consider the pair
\begin{equation}
\label{Phistar}
g_{\alpha\beta} = g^*_{\alpha\beta} \qquad \Phi^* = h^{\alpha\beta}g^*_{\alpha\beta}
\quad W_{\Phi^*}=-1
\end{equation}
If $g^*$ satisfies the Denominator Principle of \sref{denomprin} so will $\Phi^*$,
and vice versa. Let us denote the corresponding Weyl connection by $\chi^*_\alpha$.
Now consider the pair of conformal gauge metrics related by a Weyl-scaling
\begin{equation}
\label{starmap}
h'_{+-} = \Phi^* h_{+-}
\end{equation}
If $h_{+-}$ is the metric of \er{cgauge2}, the metric $h'_{+-}$ according to \er{starmap}
is the metric substitute $g^*_{+-}$. The choices \er{induced}, \er{Phichoice}, 
\er{cgauge1} and \er{cgauge2} are specific realisations of this general scheme. On using 
\er{chitrans} one finds
\begin{equation}
\label{chiprime}
{\chi^*}'_\alpha = 0
\end{equation}
Furthermore
\begin{equation}
\label{Gprime}
G'^\gamma_{\alpha\beta} = G^\gamma_{\alpha\beta}\quad {\check W}'^\gamma_{\alpha\beta}=0\quad {\Gamma ^{(2)}} '^\gamma_{\alpha\beta}= {\Gamma^{(1)}}^\gamma_{\alpha\beta}
\end{equation}
and
\begin{equation}
\label{Gconf2}
{\Gamma^{(2)}}^\gamma_{\alpha\beta} =0 \qquad {\check W}^\gamma_{\alpha\beta} = {\Gamma^{(1)}}^\gamma_{\alpha\beta}
\end{equation}
Now following the same strategy as in proving \er{identity} one shows that the Weyl-reparametrisation covariant derivatives of zero weight tensors are identical to the covariant
derivatives of \sref{covMet}. It is also easy to see that the way the metric factors and compensators matched in the example discussed earlier in this section continues to work even
for the general case and also for any action term considered. This establishes the complete
equivalence of the two formalisms as long as all tensors considered are of zero Weyl weight.

This equivalence continues to hold even when we construct actions with
tensors of non-zero Weyl weights. Firstly, the Weyl weight
compensators pick up an additional factor ${\Phi^*}^J$ where $J$ is
the sum of the Weyl weights of all the tensor factors. In place of
\er{identity} one has, when the tensors are of non-zero weight,
\begin{equation}
\label{identitygen}
\Pcm{D}_{\alpha_1\dots\alpha_n}~T_{\beta_1\dots\beta_n}={\Phi^*}^{-j}D_{\alpha_1\dots\alpha_n}~T_{\beta_1\dots\beta_n}
\end{equation}
This results in an exact compensation of the $j$-dependent factors and one ends up with the
equality of the action terms (modulo irrelevant constant factors) just as in the earlier
case of the construction with zero weight tensors.
\section{Systematic Construction of Effective Actions in Conformal Gauge}
\label{highercovterms}
We have shown how to construct manifestly covariant action terms in
both the approaches in sections \sref{covacts1} and
\sref{covacts2}. At a classical level this is all that is required. At
a quantum level, one has to work with gauge-fixed actions. As long as
the symmetries are not violated through quantum corrections, any gauge
is as good as any other. We shall restrict ourselves to the conformal
gauge and discuss the procedure for a systematic construction of
effective actions.  Nevertheless, often it is instructive to work in
different gauges both because of technical simplicity as well as for
demonstration of gauge invariance.

As we have already demonstrated the complete equivalence of the
conformal gauges of the two approaches in \sref{equiv} we shall use
the form of the results of section \sref{conf1}; one could equally
well have used section \sref{conf2}.

The systematic construction of effective action terms that are
manifestly covariant under conformal transformations proceeds more or
less along the lines of what has already been presented in
\cite{HDPM2} with some improvements suggested in \cite{Drum2}, but
with some very important differences which we address here. Before
that, we draw attention to the fact that these earlier methods were
based on using skeletal forms which were ordinary derivatives of
$X^\mu$. Because of this the transformation laws that left these
actions invariant had to be discovered each time, and by trial and
error. Our constructions in this paper now allow the skeletal forms to
be built out of covariant derivatives and because of this, invariance
of the action terms is guaranteed.

As before the method of construction involves stringing together
covariant derivatives of $X^\mu$ duly contracted with target space
invariant tensors $\eta_{\mu\nu}, \epsilon_{\mu_1\dots\mu_D}$ and then
rendered into $(1,1)$ worldsheet tensors by dividing with appropriate
powers of $L$. As before, terms proportional to the constraints and
their derivatives are dropped. Covariantly this amounts to dropping
terms proportional to $D_{\pm\pm}X\cdot D_\pm X$ and their covariant
derivatives.

The main difference from what was presented in \cite{HDPM2,Drum2}
comes in the treatment of terms proportional to EOM and its
derivatives. There they were simply dropped. As shown in detail in
\cite{HDPM2} dropping such terms amounts to a field redefinition which
can affect the transformation laws as well as the measure (in the path
integral approach). The covariant calculi presented here are based on
the fixed form of transformation laws \er{urtrans}. Therefore in the
systematic construction of terms such EOM terms can not be dropped.

Hence mixed covariant derivative terms (in the sense of having both
$+$ and $-$ indices) have to be considered in the general construction
in contrast to \cite{HDPM2,Drum2}. Even apart from the EOM issue,
\er{explicit} shows that not all mixed covariant derivatives, unlike
mixed ordinary derivatives, are proportional to EOM.

It was shown in \cite{HDPM2} that as long as one is interested in
terms up to order $R^{-3}$, such field redefinitions can be safely
carried out without worrying about the invariant measure or the
Jacobians for transformation.  The transformation laws, however, have
to be modified. A practical way out of the latter is to first work out
the full equations of motion and the full stress tensor for covariant
actions constructed by our covariant calculus and then express these
in terms of the new fields.  Even when working with action term of
higher than $R^{-3}$ order, it may prove desirable from a
calculational point of view to drop such EOM terms and carry out the
concommitant changes. The details depend on the particular case at
hand.

In the next two subsections we show how this systematic method may
be applied at the level of the PS action terms as well as the
Drummond terms at order $R^{-6}$.

As we shall see, the integrand of the PS term does not appear at
all in the covariant formulation.  In fact, it has to be treated and 
understood in a different way.
The PS term is of course essential to
`adjust' the central charge of the theory; without the PS term the
effecive string construction is not consistent outside the usual
critical dimension. 
We discuss this peculiar situation and the impossibility of 
covariantising the integrand of the PS term in section \sref{impossible}.
However, as noted there and as already known from earlier works, the {\emph action}
represented by the PS term is indeed conformally invariant.

\subsection{Attempts at covariantising the PS Terms}
In this subsection we make an attempt at covariantising the integrand
of the PS term. As discussed at length in \cite{HDPM2} there are two
(in particular) equivalent forms for the PS term that differ by total
derivative and EOM terms. These are
\begin{equation}
\label{psterm1}
I_{PS}^{(1)}=\frac{1}{L}\del_+^2~X\cdot\del_-^2~X
\end{equation}
and
\begin{equation}
\label{psterm2}
I_{PS}^{(2)}=\frac1{L^2}\del_+^2~X\cdot\del_-X~\del_-^2~X\cdot\del_+~X
.\end{equation}
This second expression is the form given in \cite{PS}, appearing in
\er{PSaction}, and we generally refer to it as ``the PS term''.

Let us consider the first of these. 
The obvious conformal-gauge candidate for this is 
\begin{equation}
\label{confpscand}
I_{PSConf}^{(1)}= L^{-1}~D_{++}~X\cdot D_{--}~X 
.\end{equation}

On using \er{explicit} \er{confpscand} can be expanded as
\begin{eqnarray}
\label{prel}
&& \del_+^2X\cdot\del_-^2X - L^{-1}\del_+L\del_-L  \nn\\
&+&  \frac{\del_+L\del_-X\cdot \del_{+-}X+\del_-L\del_+X\cdot\del_{+-}X}{L}
.\end{eqnarray}
On recalling the following identity from \cite{HDPM2}
\begin{eqnarray}
& &\frac{\partial_+^2 X\cdot\partial_-^2 X}{L} = 
\frac {\partial_+^2 X\cdot\partial_- X
\partial_-^2 X\cdot\partial_+ X}{L^2}\nonumber\\
&+&\frac {\partial_{+-}X\cdot\partial_{+-}X}{L}
- \frac{\partial_{+-}X\cdot\partial_+ X
\partial_{+-}X\cdot\partial_-X}{L^2}\nonumber\\
&+&\partial_-\big(\frac{\partial_+^2X\cdot\partial_-X}{L}\big)
-\partial_+\big(\frac{\partial_{+-}X\cdot\partial_-X}{L}\big)
\end{eqnarray}
we see that
\begin{equation}
\label{prelim}
L^{-1}D_{++}X\cdot D_{--}X = \textup{Total Derivative} + \textup{EOM}
\end{equation}

Thus though \er{confpscand} appears to be a candidate for covariant
form of \er{psterm1} it ends up being a linear
combination of total derivative terms and EOM. Through a more tedious
calculation it can be shown that the second term \er{psterm2} meets
the same fate. In fact, using the systematic procedure for
constructing actions discussed above, it is easily seen that it is not
possible to write any covariant term reproducing the PS terms. A clue
to this `anomalous' behaviour is already present in \er{psanom}.  We
shall prove this impossibility in a different and more fundamental way
in section \sref{impossible}.

\subsection{Covariantising the Drummond Terms}
\label{covIdrum}
Before proceeding, we make a few statements on terms proportional to
EOM. At this order one has to explicitly verify whether EOM terms can
be dropped or not and they can not be dropped generically. However, in
\cite{Drum} EOM terms were dropped in arriving at \er{dterms1}. Thus a
comparison can only be made if we examine the terms modulo EOM, but
otherwise we emphasise that the general construction proposed in this
paper is the more legitimate.

Let us start with \er{cov1ex1} and \er{cov1ex2}. It is easy to work out 
these expressions in the conformal gauge \er{cgauge1}:
\begin{eqnarray}
{\cal M}_1 &=& 2 \frac{D_{++}X\cdot D_{++}X~D_{--}X\cdot D_{--}X}{L^3} \nn\\
&+& 2\frac{(D_{++}X\cdot D_{--}X)^2}{L^3}
\end{eqnarray}
\begin{equation}
{\cal M}_2 = \frac{4}{L^3}(D_{++}X\cdot D_{--}X)^2
\end{equation}
We consider the particular combination
\begin{equation}
{\cal M}_1-\frac{{\cal M}_2}{2} = \frac{2}{L^3}(D_{++}X\cdot D_{++}X)(D_{--}X\cdot D_{--}X)
,\end{equation}
and it is easy to show that, modulo terms that are leading-order
constraints and their derivatives, this is just $M_1$. To
understand ${\cal M}_2$ let us display \er{prel} slightly differently
as
\begin{equation}
\label{prel3}
L^{-1}D_{++}X\cdot D_{--}X = \del_-(L^{-1}\del_+^2X\cdot\del_-X) + \textup{EOM}
\end{equation}
Then it follows that
\begin{eqnarray}
{\cal M}_2 &=& L^{-1}[L^{-1}\del_+^2X\cdot\del_-^2X-L^{-2}\del_-L~\del_+^2X\cdot\del_-X]^2\nn\\
&=& M_2-2M_3 +M_4
\end{eqnarray}
This way we are able to obtain two independent linear combinations of \er{dterms1}.
It can be shown, through straightforward but tedious algebra, that the covariant
calculus can not produce any other combinations. The obvious approach to covariantising
the \er{dterms1} by replacing ordinary derivatives by covariant derivatives only
produces, apart from these combinations, EOM and derivatives, constraints and their derivatives, and
total derivatives. This is completely analogous to the situation with PS terms 
discussed in the previous secion.

The present formalism, while representing quite a general way of
formulating covariance, is thus extremely restrictive. 
The only possible gauge-fixed action to $R^{-6}$ order is, up to irrelevant terms,
\begin{equation}
\label{covallowed}
\int\frac{\td^2 \sigma}{4\pi} \bigg( \frac{L}{a^2} + \beta \frac{\del_+ L\del_- L}{L^2}  + \beta_1 ( M_2 - 2M_3 +M_4 ) +\beta_2 M_1 \bigg)
.\end{equation}
Drummond's
terms appear, but only in these particular combinations.  
This is one of the main results of the present paper; which we
now emphasise. The effective string action has only three parameters
$\beta,\beta_1,\beta_2$ (in addition to the string tension $a^2$ which merely sets
a physical scale) at order $R^{-6}$.
Of these the second term, which corresponds to the PS term, does not transform covariantly.
Nevertheless, the PS action is invariant under the conformal transformations.
Of the three parameters, the leading order analysis as given by PS already fixes
$\beta$ leaving only two free parameters $\beta_1,\beta_2$. It would be interesting to see if higher
order analysis would further fix some of the remaining parameters.

\section{Impossibility of Covariantising the PS Integrand}
\label{impossible}

We shall prove that for WZNW effective actions for a conformal anomaly
in two dimensions defined by
\begin{equation}
\label{WZNW}
\delta_\lambda S_{WZNW} = \int \td^2\xi \lambda(\xi)\sqrt{g}R(\xi)
\end{equation}
the integrand of $S_{WZNW}$ can never be manifestly covariant under coordinate
transformations. 

Here, $R(\xi)$ is the Ricci scalar.  The action
proposed by Polyakov \cite{polyagrav} in the context of two dimensional quantum
gravity,
\begin{equation}
S_{\textup{Polya}} = \int R \frac{1}{\nabla^2} R
\end{equation}
is such a WNZW action.

Written out explicitly,
\begin{equation}
\label{pol}
S_{\textup{Polya}} = \int \td^2 x \sqrt{g(x)} R(x) (\frac{1}{\nabla^2} R)(x)
\end{equation}
where $R(x)$ is the scalar curvature in two dimensions. 
The value of the integrand in the conformal gauge of \er{cgauge1} is
\begin{eqnarray}
I_{\textup{Polya}} &=& (\del_+\del_-~\ln L)\cdot\frac{1}{\del_+\del_-}\cdot(\del_+\del_- \ln L)\nn\\
          &=& (\del_+\del_-~\ln L)\ln L
\end{eqnarray}
This is the same as the integrand $L_2$ of \er{defL2} upto total derivative terms.
The variation of this under conformal transformations \er{urtrans} is
\begin{equation}
\label{polvar}
\delta I_{\textup{Polya}} 
= \del_-(\epsilon^-~I_{\textup{Polya}})+\del_-\epsilon^-~\del_+\del_-\ln L
\end{equation}
Since $\del_+\epsilon ^-=0$, it follows that the Polyakov action is indeed invariant under
conformal transformations.

The Weyl scalings are defined by
\begin{equation}
\delta_{\lambda(\xi)}~g_{\alpha\beta} = \lambda(\xi)~g_{\alpha\beta}
\end{equation}
and the infintesimal coordinate transformations are given by
\begin{equation}
\delta_\epsilon~g_{\alpha\beta} = \epsilon^\gamma\del_\gamma g_{\alpha\beta}+\del_\alpha\epsilon^\gamma
g_{\gamma\beta}+\del_\beta\epsilon^\gamma g_{\alpha\gamma}
\end{equation}
Now we look for a combination of Weyl scaling and coordinate
transformation that leaves the form of the metric unchanged;
\begin{equation}
\label{totdef}
\bar\lambda g_{\alpha\beta} = 
 \epsilon^\gamma\del_\gamma g_{\alpha\beta}+\del_\alpha\epsilon^\gamma
g_{\gamma\beta}+\del_\beta\epsilon^\gamma g_{\alpha\gamma}
.\end{equation}
The strategy is to consider
\begin{equation}
\delta_{tot} = \delta_{\bar\lambda}-\delta_\epsilon
,\end{equation}
and by construction
\begin{equation}
\label{total}
\delta_{tot} \tilde{\cal L} = 0
.\end{equation}
Although we are talking about the same transformations we discussed in
section \sref{review}, it is worth emphasising that here \er{total}
does not say anything about the invariance or lack of invariance of
the action under any of the said transformations.  On the other hand,
we have, from \er{WZNW}
\begin{equation}
\label{weylvar}
\delta_{\textup{Weyl}}(\bar\lambda) \tilde{\cal L} = \bar\lambda \sqrt{g} R(\xi) + \del_+ X^+(g,\bar\lambda) + \del_-X^-(g,\bar\lambda)
\end{equation}
It should be noted that the dependence of $X^\pm$ is explicitly on
$\bar\lambda$ and its derivatives.

For the remainder we work explicitly in the conformal gauge of
\er{cgauge1} and without loss of generality, restrict our attention to
only $\xi^-$ diffeomorphisms ($\epsilon^+=0$). In this case
\er{totdef} reads
\begin{equation}
\label{totcgauge}
\del_+\epsilon^-=0;~~~~\bar\lambda = \del_-\epsilon^-+\epsilon^-~\del_-\ln L
\end{equation}
Using \er{ricciscalar} and \er{totcgauge} we rewrite \er{weylvar} as
\begin{eqnarray}
\label{explicit2}
\delta_{\textup{Weyl}}(\bar\lambda)~\tilde{\cal L} 
&=&\del_-\{-2\epsilon^-\del_{+-}\ln L+X^-\} \\
&+&\del_+\{2\epsilon^-\del_{--}\ln L -\epsilon^-(\del_-\ln L)^2+X^+\} \nn
\end{eqnarray}
Now, this must equal $\delta(\epsilon^-)\tilde{\cal L}$. If
$\tilde{\cal L}$ were transforming as a scalar density, the $\del_+$
terms in the last line of \er{explicit2} must vanish identically. This
can happen only if
\begin{equation}
\label{xplus}
\del_+X^+ = \del_+\{\epsilon^-(-2\del_{--}\ln L+(\del_-\ln L)^2)\}
\end{equation}
As we have already noted $X^+$ must have an explicit dependence on
$\bar\lambda$ and its derivatives. Part of \er{xplus} can indeed be
cast into this form (which part can be so cast is not unique);
\begin{eqnarray}
\del_+X^+ &=&  \del_+\{-2\del_-(\epsilon^-\del_-\ln L+\del_-\epsilon^-) \\
&+&(\epsilon^-\del_-\ln L+\del_-\epsilon^-)\del_-\ln L+\del_-\epsilon^-\del_-\ln L\} \nn
.\end{eqnarray}
The last term $\del_-\epsilon^-\cdot\del_{+-}\ln L$ makes it evident
that $\tilde{\cal L}$ fails to be a scalar density. It fails by
precisely the same term as obtained through explicit calculations with
$I_{\textup{Polya}}$, which we have seen in \er{polvar}. Nevertheless, $S_{Polya}$
is invariant as shown above.

\section{Conclusions}
We have given in this paper a vastly simplified approach to the theory
of effective strings in comparison with the PS formalism. The
essential simplification is that the transformation law is always the
same as the standard transformation law for free bosonic string
action. The transformation law does not have to be tuned to the
action. In the conformal gauge, these represent conformal
transformations exactly, and not approximately as in the case of the
PS transformation law( only to order $R^{-3}$ to be precise).

Consequently every action constructed by our covariant calculus, and
in particular $S_{2}$ of \er{action2}, is in principle valid to \emph
{all orders} in $R^{-1}$. Whether phenomenologically any action or
combination of actions is correct or not is a different issue.

A further consequence of this covariantisation is the restriction of
the effective action to order $R^{-6}$ to include only two free
parameters $\beta_1$ and $\beta_2$.  Our result for the complete
effective action to this order, from our first approach, is the
truncation to order $R^{-6}$ of
\begin{eqnarray}
\label{covallowed2}
S &=&\frac{\beta}{4\pi} S_{\textup{Polya}}+\int\frac{\td^2\sigma}{4\pi}\big[\frac{\sqrt{g}}{a^2}+ 
\beta_1 X^{;\alpha\beta}\cdot X^{;\delta\gamma} X_{;\alpha\beta}\cdot X_{;\delta\gamma} \nn\\
&+& \beta_2 X^{;\alpha\beta}\cdot X_{;\alpha\beta}X^{;\gamma\delta} \cdot X_{;\gamma\delta}
\big] ,\nn\\
\end{eqnarray}
The conformal gauge expression for \er{covallowed2} is given by
\er{covallowed} and that entire action is by construction
\emph{exactly} conformally invariant under the transformation law
\eqref{urtrans}. Both the approaches yield the same actions in the
conformal gauge.  As already emphasised before, one can use the entire
\er{covallowed}, without truncation, if one so wishes.  Further terms
presumably begin to appear at $\Pcm{O}(R^{-8})$.

Finally, it is worth emphasising that of course without a covariant
formulation, one can only identify the relevant terms which may be
included in the action, up to a given order, and adorn these terms
with coefficients which are then the parameters of the theory. These
must then be determined phenomenologically (using lattice QCD, for
example). In contrast, in our covariant construction, in general the
number of new free parameters introduced at each order in $1/R$ is
fewer than the number which would obtain given such independent
insertion of all relevant terms. This highly desirable reduction in
parameters is also a conclusion which could be subject to verification
in simulations or experiments.

\section*{acknowledgement}
NDH acknowledges the award of the DAE Raja Ramanna Fellowship enabling
his stay at IISc.


\begin{thebibliography}{99}
\bibitem{PS} Joseph Polchinski and Andrew Strominger, \emph{Effective String Theory}, PRL 67, 1681 (1991).
\bibitem{Drum} J.~M.~Drummond, \emph{Universal Subleading Spectrum of Effective String Theory}, \texttt{hep-th/0411017}.
\bibitem{HDPM2} N.~D.~Hari Dass and Peter Matlock, \emph{Field Definitions, Spectrum and Universality in Effective String Theories}, \texttt{hep-th/0612291}.
\bibitem{kretsch} E.~Kretschmann, Ann. Phys. Leipzig. {\bf 53}, 575(1917).
\bibitem{Drum2} J.~M.~Drummond, \texttt{hep-th/0608109v1}.
\bibitem{polyagrav} A.M. Polyakov, \emph{Quantum Gravity in Two Dimensions}, Mod.Phys.letts. {\bf A2} 893(1987).
\end{thebibliography}
\end{document}